Crystallography of the martensitic transformation between Ni$_2$In-type hexagonal and TiNiSi-type orthorhombic phases

Tingting Zhang, Yuanyuan Gong*, Bin Wang, Dongyu Cen, Feng Xu*

MIIT Key Laboratory of Advanced Metallic and Intermetallic Materials Technology, Nanjing University of Science and Technology, Nanjing 210094, China

*corresponding author: gyy@njust.edu.cn; xufeng@njust.edu.cn

**Abstract**

MnMX (M = Co or Ni, X = Si or Ge) alloys, experiencing structural transformation between Ni$_2$In-type hexagonal and TiNiSi-type orthorhombic phases, attract considerable attention due to their potential applications as room-temperature solid refrigerants. Although lots of studies have been carried out on how to tune this transformation and obtain large entropy change in a wide temperature region, the crystallography of this martensitic transformation is still unknown. The biggest obstacle for crystallography investigation is to obtain a bulk sample, in which hexagonal and orthorhombic phases coexist, because the MnMX alloys will fragment into powders after experiencing the transformation. For this reason, we carefully tune the transformation temperature to be slightly below 300 K. In that case, a bulk sample with small amounts of orthorhombic phases distributed in hexagonal matrix is obtained. Most importantly, there are no cracks between the two phases. It facilities us to investigate the microstructure using electron microscope. The obtained results indicate that the orientation relationship between hexagonal and orthorhombic structures is $[4\bar{2}\bar{2}3]_h // [120]_o$ & $(01\bar{1}0)_h // (001)_o$ and the habit plane is $\{\bar{2}113.26\}_h$. WLR theory is also adopted to calculate the habit plane. The calculated result agrees well with the measured one. Our work reveals the crystallography of hexagonal-orthorhombic transformation for the first time and is helpful for understanding the transformation-associated physical effects in MnMX alloys.

## 1. Introduction

MnCoSi, MnCoGe, MnNiSi and MnNiGe (denoted as MnMX (M=Co or Ni, X=Ge or Si) hereafter) are four typical compounds, which experience reversible structural transformation between high-temperature $Ni_2In$-type hexagonal (space group P63/mmc, 194) and low-temperature TiNiSi-type orthorhombic (space group Pnma, 62) phases [1-14]. This structural transformation displays the following advantages: (i) be accompanied by large entropy change, (ii) can be induced by multiple sources, such as heat, pressure as well as magnetic field, (iii) the transformation temperature can be tuned in a large temperature region covering room temperature by doping [3-29]. These features make MnMX-based alloys become new room-temperature solid refrigerants [5-7,20,30].

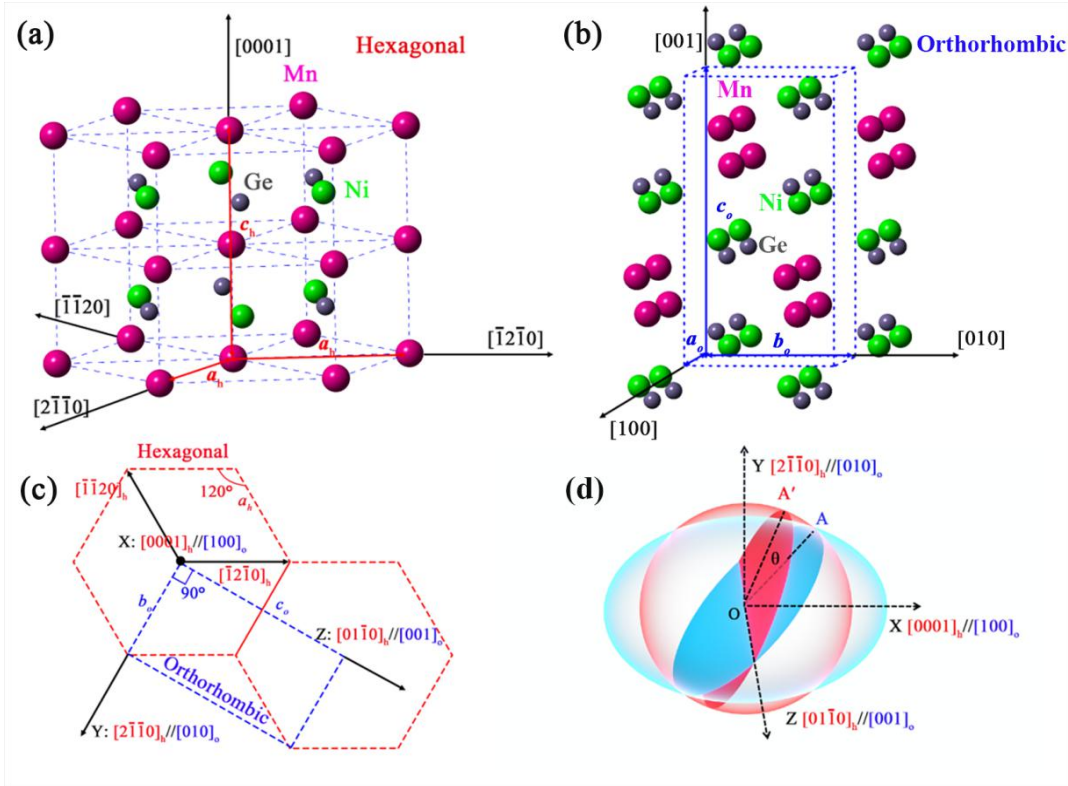

**Fig. 1.** Crystal structures of (a) hexagonal and (b) orthorhombic MnNiGe. (c) Schematic diagram of the relationship between the hexagonal and orthorhombic structures with the following orientation relationship: $[0001]_h // [100]_o$, $[2\bar{1}\bar{1}0]_h // [010]_o$ & $[01\bar{1}0]_h // [001]_o$. In (c), $[0001]_h$ and $[100]_o$ are vertical to the paper. (d) A sphere→ellipsoid transformation model, in which the red sphere and blue ellipsoid express as hexagonal and orthorhombic phases,

respectively. After experiencing the strain in Formula (1), the red sphere transforms into the bulk ellipsoid.

Take MnNiGe as an example (see Fig. 1a and b). The hexagonal structure is consisted of the honeycomb layers stitched up by Mn chains along $c_h$ axis [31]. Covalent bonds form between Ni and Ge atoms, while Mn is relative isolated, resulting large magnetic moment [1,32-36]. During the structural transformation, all the atoms show changes in their coordination and the atomic chain along $a_o$ axis becomes a zigzag arrangement in orthorhombic structure [34]. (The subscripts of h and o indicate hexagonal and orthorhombic structures, respectively.) This structural transformation is diffusionless and firstly referred to as martensitic transformation by koyama in 2004 [8].

As a martensitic transformation, there should be a habit plane, which is also known as an invariant plane or the interface between martensite and austenite. In this plane, vectors don't strain and rotate during the martensitic transformation [37]. For the hexagonal-orthorhombic transformation in MnMX alloys, although lots of studies have been carried out on how to tune the transformation temperature and obtain large entropy change in a wide temperature region [3-5,9-10,16-18,21-22,30-31], the habit plane as well as the crystallography of this transformation are still unclear. Till now, we only know that, as most previous studies mentioned, the hexagonal lattice parameters ($c_h$ and $a_h$) are related to the orthorhombic parameters ($a_o$, $b_o$ and $c_o$) in the following manner: $c_h \to a_o, a_h \to b_o$ and $\sqrt{3}a_h \to c_o$ [1,10-14,16,18-19,23]. To clearly show this manner, a XYZ rectangular coordinate is established with X= $[0001]_h // [100]_o$, Y= $[2\bar{1}\bar{1}0]_h // [010]_o$ & Z= $[01\bar{1}0]_h // [001]_o$ (see Fig. 1c). Here, $c_h$ (or $[0001]_h$) and $a_o$ (or $[100]_o$) are vertical to the paper for simplicity. It should be noted that this rectangular coordinate can also be established with X= $[0001]_h // [100]_o$, Y= $[\bar{1}2\bar{1}0]_h // [010]_o$ & Z= $[\bar{1}010]_h // [001]_o$ or X= $[0001]_h // [100]_o$, Y= $[\bar{1}\bar{1}20]_h // [010]_o$ & Z= $[1\bar{1}00]_h // [001]_o$ because $[2\bar{1}\bar{1}0]_h$, $[\bar{1}2\bar{1}0]_h$ and $[\bar{1}\bar{1}20]_h$ are crystallographically equivalent. Here, we just take the situation shown in Fig. 1c as an example. Since the transformation from hexagonal to orthorhombic phases displays a shrinkage along Y axis, an expansion along X axis and a negligible variation along Z axis (see Table 1, discussed later), the deformation that connects the hexagonal and orthorhombic structures is expressed as

$$\begin{bmatrix} X_2 \\ Y_2 \\ Z_2 \end{bmatrix}_o = \varepsilon \begin{bmatrix} X_1 \\ Y_1 \\ Z_1 \end{bmatrix}_h \quad \text{and} \quad \varepsilon = \begin{bmatrix} \frac{a_o}{c_h} & & \\ & \frac{b_o}{a_h} & \\ & & \frac{c_o}{\sqrt{3}a_h} \end{bmatrix} \quad (1)$$

in the rectangular coordinate shown in Fig. 1c. The matrix ε is the deformation matrix, which is similar to the famous Bain strain [37].

However, we should mention that ε represents the lattice deformation from hexagonal to orthorhombic structures, but it doesn't mean that the hexagonal and orthorhombic structures obey the following orientation relationship: $[0001]_h // [100]_o$, $[2\bar{1}\bar{1}0]_h // [010]_o$ & $[01\bar{1}0]_h // [001]_o$. Because, if so, there will be no habit plane according to the phenomenological crystallographic theory of martensitic transformation [37]. The reason is summarized as follows in brief. According to the sphere→ellipsoid transformation model [37], the parent structure (hexagonal structure) is represented as a red sphere (see Fig. 1d). After the deformation strain ε, it transforms into the blue ellipsoid with X expanded, Y shrunken and Z unchanged. The sphere and ellipsoid meet at A point in the XY plane. The vector OA and Z axis are both in the blue plane. The initial position of the blue plane is marked in red. According to Fig. 1d, it can be found that the blue plane is the only plane in which the vector (i.e. OA) is not strained during the transformation, but rotated by θ from its initial position (OA'). So, there are no unstrained and unrotated vectors, and of course no habit plane.

The above-mentioned analysis proves that the strain ε can just be used to represent the lattice deformation between hexagonal and orthorhombic structures, but it doesn't reflect the real orientation relationship since this orientation relationship ( $[0001]_h // [100]_o$, $[2\bar{1}\bar{1}0]_h // [010]_o$ & $[01\bar{1}0]_h // [001]_o$ ) can't fulfill the basic requirements of bringing about a martensitic transformation. Now, questions arise: what is the real orientation relationship between the orthorhombic and hexagonal structures? And what is the habit plane?

To answer the questions, carrying out microstructure investigation on an area, where orthorhombic and hexagonal phases coexist, is a promising way. So, one can choose one sample with a room-temperature structural transformation and measure the real orientation relationship using i.e. transmission electron microscope or electron backscattered diffraction. However, the situation is more complicated than thought. In the studies on MnCoGe, MnNiGe, MnNiSi as well as MnCoSi, researchers all found that the sample will fragment after experiencing the transformation because of the large lattice parameter change accompanied by the transformation [8-14,16-19,23,39-41]. It means that the bulk sample will fragment once the transformation occurs and the interface between the two phases becomes hard to be found. It seems impede the crystallography investigation. Also for this reason, microstructure, especially the microstructure of a hexagonal/orthorhombic-coexisted sample, has seldom been showed in most previous studies.

In this paper, we take MnNiGe as an example and tune the transformation temperature to be slightly below 300 K. Amazingly, some fine laths in orthorhombic structure are observed in the hexagonal matrix. Most importantly, there are no cracks between orthorhombic and hexagonal phases. It provides us an opportunity to reveal the orientation relationship and the habit plane. As we known, it is the first time to provide the microstructure image of a orthorhombic/hexagonal-coexisted sample and also the first time to reveal the orientation relationship as well as the habit plane in MnMX alloys.

**2. Experimental**

Polycrystalline samples with nominal compositions of MnNi$_{0.84}$V$_{0.16}$Ge and MnNi$_{0.83}$V$_{0.17}$Ge were prepared by arc-melting high-purity raw materials under argon atmosphere for three times. The as-cast ingots were annealed in evacuated quartz tubes at 1073 K for five days, followed by quenching into water. The crystal structures of the samples were identified by powder X-ray diffraction (XRD) with Cu-Ka radiation using Bruker D8 Advance at room temperature. The structure refinement of the XRD patterns was performed using Fullprofs implementation of the Rietveld refinement method. Magnetic measurements were performed using a cryogen-free physical property measurement system (PPMS Dynacool™) with a vibrating sample magnetometer from Quantum Design. A bulk sample with a mass of ~20 mg is used for magnetic measurement. During the magnetic measurement, the sample is firstly cooled to 100 K under zero field then measured in one heating-cooling cycle under 0.1 T. The microstructure was investigated using a FEI Quanta 250F scanning electron microscope (SEM) and optical microscope. Before the observation, the surface was mechanical polished then corroded by 60% $HNO_3$ for 10 seconds at room temperature. Elemental mapping was taken by energy-dispersive spectroscopy (EDS) in the SEM. Conventional transmission electron microscope (TEM) measurement was performed on an FEI Tecnai G2 F20 microscope. The high-resolution transmission electron microscope (HRTEM) and scanning transmission electron microscope (HRSTEM) observations were carried out using an FEI Titan Themis microscope. The sample for TEM measurement was prepared by ion beam milling technique.

## 3. Results and discussion

3.1 Coexistence of orthorhombic and hexagonal phases

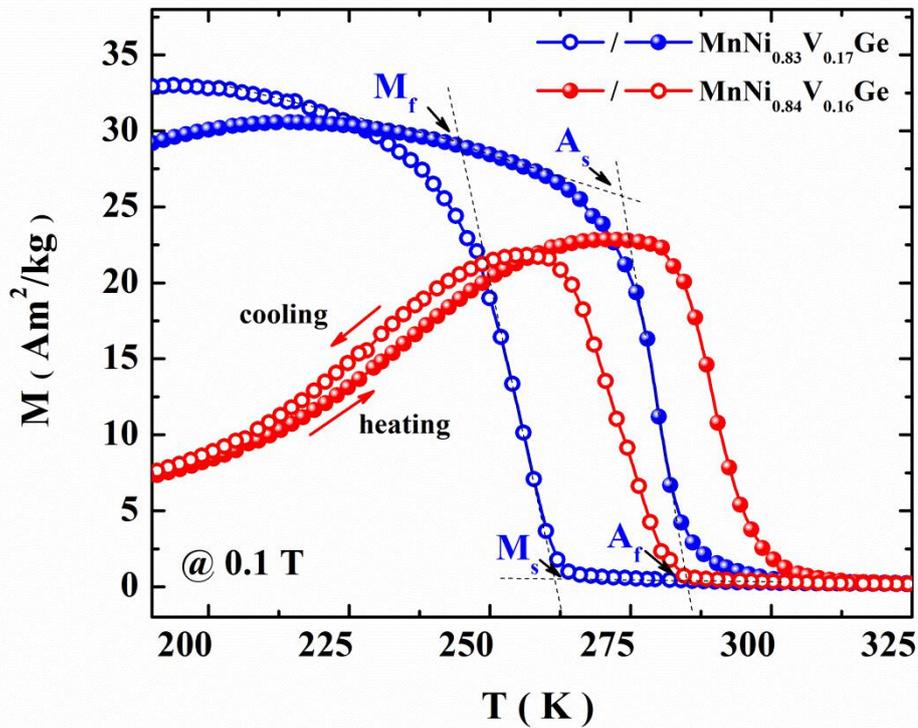

**Fig. 2.** The M-T curves of MnNi$_{0.84}$V$_{0.16}$Ge and MnNi$_{0.83}$V$_{0.17}$Ge alloys.

Stoichiometric MnNiGe alloy undergoes the structural transformation from hexagonal to orthorhombic phases at ~470 K during cooling [1]. The transformation temperature decreases with substituting V for Ni and be close to 300 K when the V-content reaches 16-17% [42]. For MnNi$_{0.84}$V$_{0.16}$Ge or MnNi$_{0.83}$V$_{0.17}$Ge alloy, the hexagonal phase is paramagnetic while the orthorhombic phase is ferromagnetic-like [42]. It facilitates us to estimate the structural transformation temperature based on the temperature dependences of magnetization (M-T) curves. The M-T curves for MnNi$_{0.84}$V$_{0.16}$Ge and MnNi$_{0.83}$V$_{0.17}$Ge alloys under a magnetic field of 0.1 T are shown in Fig. 2. The sharp magnetization increase during cooling (decrease during heating) indicates the transformation from hexagonal to orthorhombic (orthorhombic to hexagonal) phases. Thermal hysteresis, expressing as the misalignment between heating and cooling curves, proves the first-order nature of this transformation. The transformation in MnNi$_{0.84}$V$_{0.16}$Ge alloy occurs at a higher temperature than that in MnNi$_{0.83}$V$_{0.17}$Ge alloy. The martensitic transformation start temperature ($M_s$), martensitic transformation finish temperature ($M_f$), austenitic transformation start temperature ($A_s$) and austenitic transformation finish temperature ($A_f$) are 282.6 (261.5) K, 261.7 (244.8) K, 282.2 (273.4) K and 297.3 (285.1) K for MnNi$_{0.84}$V$_{0.16}$Ge (MnNi$_{0.83}$V$_{0.17}$Ge), respectively.

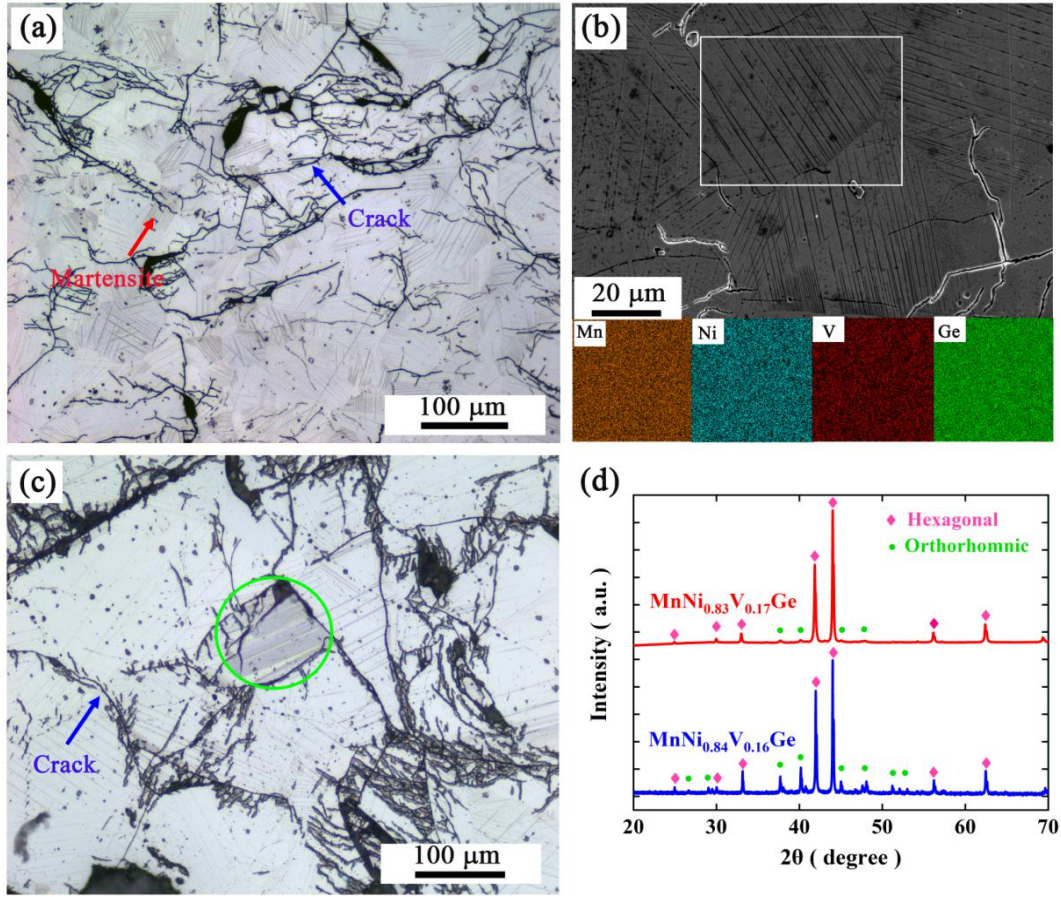

**Fig. 3.** Optical images of (a) MnNi$_{0.83}$V$_{0.17}$Ge and (c) MnNi$_{0.84}$V$_{0.16}$Ge. (b) SEM image of MnNi$_{0.83}$V$_{0.17}$Ge. (d) room-temperature XRD patterns of MnNi$_{0.83}$V$_{0.17}$Ge and MnNi$_{0.84}$V$_{0.16}$Ge.

The morphology of MnNi$_{0.83}$V$_{0.17}$Ge alloy taken from optical microscope is shown in Fig. 3a. One can found some curved cracks, which may arise from anisotropic thermal expansion of hexagonal phase. Besides these cracks, lots of straight and thin laths, which are parallel to one another, can be clearly observed. These laths seem start from the grain boundary then develop into the grain. The observed laths show the same feature with the martensite laths observed in NiMnGa, TiNi and some other compounds [43-46]. SEM image and EDS mapping for this surface are shown in Fig. 3b. There are no composition differences between the laths and matrix, indicating that the observed laths are not precipitates induced by V doping. Fig. 3d further shows the XRD patterns of MnNi$_{0.83}$V$_{0.17}$Ge alloy. Besides the main peaks of hexagonal phase, some hints of orthorhombic phase can be found. According to the results mentioned above, it is reasonable to believe that the observed laths in MnNi$_{0.83}$V$_{0.17}$Ge alloy are orthorhombic phase distributed in hexagonal matrix. With decreasing V-content to 0.16, the martensitic transformation temperature will become closer to room temperature and therefore the amount of orthorhombic phase will increase. The XRD patterns of MnNi$_{0.84}$V$_{0.16}$Ge (Fig. 3d) show enhanced peaks of orthorhombic phase. In the optical image of MnNi$_{0.84}$V$_{0.16}$Ge alloy (Fig. 3c), the narrow laths become a relative larger area with orthorhombic structure (green circle in Fig. 3c). These observations further confirm that the observed laths are in orthorhombic structure. Actually, similar laths have also been observed by Quintana-Nedelcos in MnCoGe-based sample [47]. But they are described as ferroelastic domains and further investigation is not carried out. Since the measured transformation temperature is slightly lower than 300 K, the observed laths can also be considered as the nucleus of orthorhombic phase.

It is well known that the MnMX (M=Co or Ni, X=Si or Ge) bulk will fragment and become powder-like after experiencing the structural transformation [8-14,19,23-27,38]. However, our optical and SEM images show that there are no cracks between the two phases. It indicates that fragmentation is not a phenomenon once the orthorhombic phase forms but should be a result of the generation of large amounts of orthorhombic phase. For MnNi$_{0.84}$V$_{0.16}$Ge and MnNi$_{0.83}$V$_{0.17}$Ge alloys, the martensitic transformation temperature is relatively lower than 300 K and only a small amount of thin martensite laths form (or nucleate) at room temperature. Therefore, the two samples haven't changed to be powder-like.

The microstructure shown in Fig 3a,c can not only be observed in the V-doped MnNiGe alloys but also in a slow-cooled Mn$_{0.88}$Fe$_{0.12}$NiGe alloy with a reported transformation temperature of ~280 K [23] (Fig. S1). It means that these thin orthorhombic laths can be easily obtained in the MnMX alloys at room temperature in the case of that the transformation temperature is slightly below 300 K.

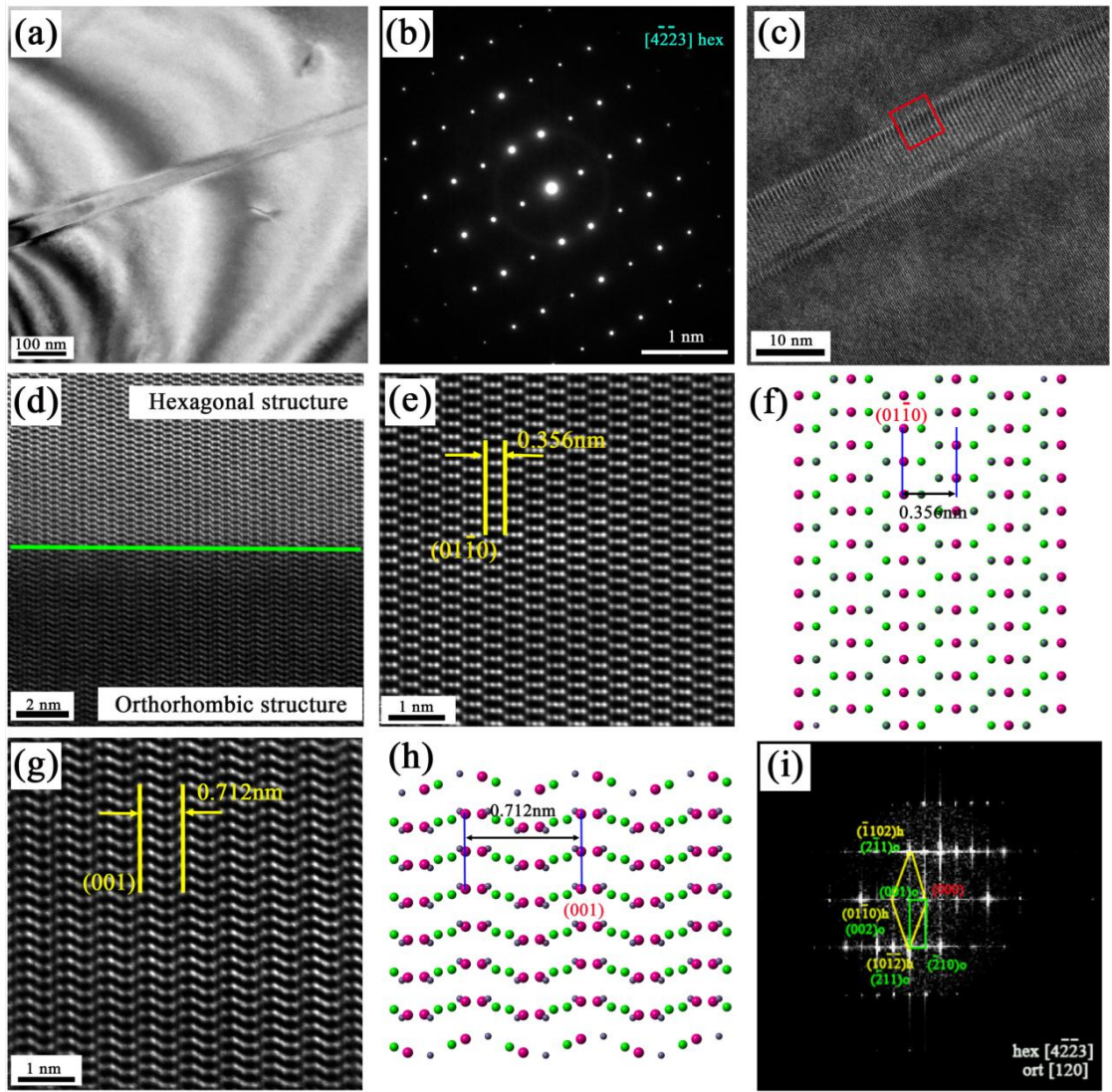

**Fig. 4.** (a) Bright-field TEM image of a foil, in which hexagonal and orthorhombic phases coexist. (b) Selected-area electron diffraction pattern of hexagonal structure. (c) HRTEM image taken from an area where one acicular orthorhombic phase exists. (d) HRSTEM image taken from the red square in (c). HRSTEM images taken from (e) hexagonal and (g) orthorhombic phases. (f) Atomic configuration of hexagonal structure with $[4\bar{2}\bar{2}3]_h$ vertical to the paper and $[01\bar{1}0]_h$ horizontal. (h) Atomic configuration of orthorhombic structure with $[120]_o$ vertical to the paper and $[001]_o$ horizontal. (i) Fast fourier transform of (d).

3.2 Orientation relationship between hexagonal and orthorhombic structures

Thanks to the observed phase coexistence in MnNi$_{0.84}$V$_{0.16}$Ge and MnNi$_{0.83}$V$_{0.17}$Ge alloys, we have an opportunity to investigate the crystallography of the structural transformation. A 86 nm-thick foil is cut by focused ion beam from MnNi$_{0.84}$V$_{0.16}$Ge alloy. The cutting plane is perpendicular to the length of the lath (see Fig. S2 in supplemental material). In the bright-field TEM image of this foil (Fig. 4a), one acicular orthorhombic phase penetrating across the whole foil can be clearly found. Selected-area electron diffraction pattern of hexagonal matrix (Fig. 4b) is consistent with $[4\bar{2}\bar{2}3]_h$ zone. We keep the electron beam parallel to this direction and carry out HRTEM and HRSTEM observations. HRTEM image taken from an area where one acicular orthorhombic phase appears is shown in Fig. 4c. It can be clearly found that the matrix and the acicular phase have different atomic arrangements. HRSTEM image taken from the red square, where the hexagonal and orthorhombic phases coexist, is displayed in Fig. 4d. The upper and lower parts of Fig. 4d are hexagonal and orthorhombic phases, respectively. The green line indicates the interface, also the habit plane. A magnified HRSTEM image taken from hexagonal phase is shown in Fig. 4e. If three adjacent atoms are connected into a short line, the atomic configuration can be obtained by periodicity arranging these short lines into a step shape. It is the same with the atomic configuration of hexagonal structure with $[4\bar{2}\bar{2}3]_h$ vertical to the paper and $[01\bar{1}0]_h$ horizontal (Fig. 4f). The $(01\bar{1}0)_h$ surface, which is vertical to the paper, is also shown in Fig. 4e,f. The interspace of $(01\bar{1}0)_h$ is 0.356 nm. This value is equal to $\sqrt{3}a_h/2$ (XRD refinement indicates that $a_h$ = 0.411 nm and $c_h$ = 0.539 nm). Magnified HRSTEM image taken from orthorhombic phase is shown in Fig. 4g. The atomic arrangement becomes wavery shape, which is the same with the atomic configuration of orthorhombic structure with $[120]_o$ vertical to the paper and $[001]_o$ horizontal (Fig. 4h). The $(001)_o$ surface, which is also vertical to the paper, is included in Fig. 4g,h. The interspace of $(001)_o$, which is the length of $c_o$ axis, is 0.712 nm. This value is the same with the value of $c_o$ obtained from XRD refinement ($a_o$=0.604 nm, $b_o$=0.378 nm and $c_o$=0.712 nm). More information about atomic occupation can be found in Table S1, S2 and Fig. S3 in supplemental material. According to the above-mentioned results, the orientation relationship between hexagonal and orthorhombic structures can be confirmed to be:

$[4\bar{2}\bar{2}3]_h // [120]_o$ & $(01\bar{1}0)_h // (001)_o$.

This orientation relationship is supported by fast fourier transform of Fig. 4d (see Fig. 4(i)). In the case of $[4\bar{2}\bar{2}3]_h // [120]_o$ & $(01\bar{1}0)_h // (001)_o$, $(\bar{1}102)_h // (2\bar{1}1)_o$ and $(10\bar{1}\bar{2})_h // (\bar{2}11)_o$ are meanwhile fulfilled. These orientation relationships of the two phases are summarized and exhibited using stereographic projection of crystal indices (Fig. S4) for clarity. In this paper, we select the low index lattice plane to express the orientation relationship, which is still $[4\bar{2}\bar{2}3]_h // [120]_o$ & $(01\bar{1}0)_h // (001)_o$.

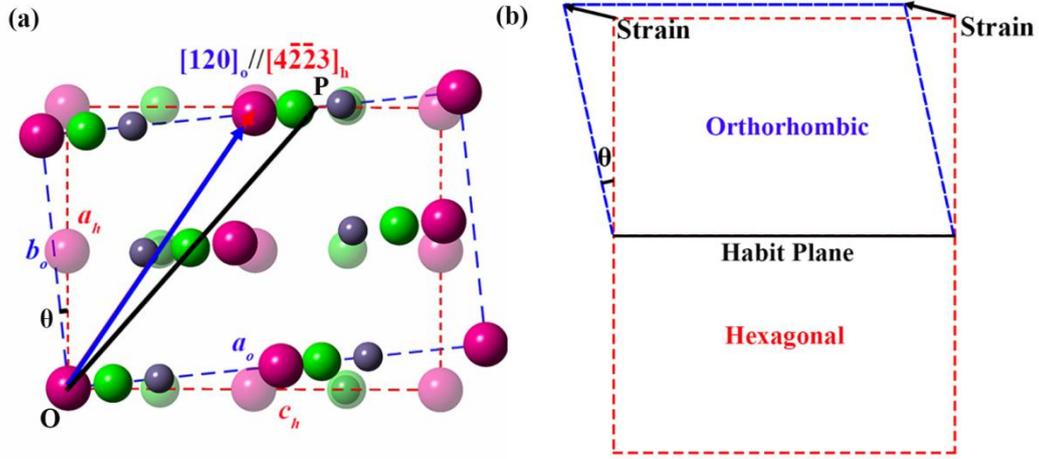

**Fig. 5.** (a) Atomic arrangement of $(001)_o$ and $(01\bar{1}0)_h$ surfaces. $[4\bar{2}\bar{2}3]_h$ is kept to be parallel to $[120]_o$. (b) Illustration of the strain, which induce the transformation from hexagonal to orthorhombic structures. In both (a) and (b), $[001]_o$ and $[01\bar{1}0]_h$ are vertical to the paper.

## 3.3 Habit plane

In the following text, we will confirm the habit plane according to the TEM results mentioned above. To confirm the habit plane, we need to find two invariant lines, which don't strain and rotate during the transformation. The two lines should also be not parallel to one another so that they can create the habit plane [37].

As shown in Fig. 4d, both $[01\bar{1}0]_h$ and $[001]_o$ are parallel to the habit plane. $[01\bar{1}0]_h // [001]_o$ agrees well with the previous opinion: $\sqrt{3}a_h \rightarrow c_o$ (see Fig. 1c in introduction). Therefore, $c_o \rightarrow \sqrt{3}a_h$ should be without length change during the transformation. According to the papers on MnMX alloys published before, we summarize the lattice parameter changes in different MnMX alloys in Table 1. It can be clearly found that the variation of $\sqrt{3}a_h \rightarrow c_o$ (or $c_o \rightarrow \sqrt{3}a_h$) is near-zero. This result confirms that $[001]_o$ and $[01\bar{1}0]_h$ are in the habit plane and $[01\bar{1}0]_h$ is just one invariant line.

Table 1 Variation of lattice parameters

| Composition | Variation | | |
|---|---|---|---|
| $Mn_{0.84}Fe_{0.16}NiGe$ [23] | $a_h \rightarrow b_o$ Δ=-9.40% | $c_h \rightarrow a_o$ Δ=12.26% | $\sqrt{3}a_h \rightarrow c_o$ Δ=0.05% |
| $Mn_{0.9}Fe_{0.2}Ni_{0.9}Ge$ [10] | $a_h \rightarrow b_o$ Δ=-8.48% | $c_h \rightarrow a_o$ Δ=12.25% | $\sqrt{3}a_h \rightarrow c_o$ Δ=0.12% |
| $MnNiGe_{0.9}Ga_{0.1}$ [11] | $a_h \rightarrow b_o$ Δ=-8.5% | $c_h \rightarrow a_o$ Δ=12% | $\sqrt{3}a_h \rightarrow c_o$ Δ=0.15% |
| $Mn_{1.07}Co_{0.92}Ge$ [8] | $a_h \rightarrow b_o$ Δ=-7.7% | $c_h \rightarrow a_o$ Δ=11% | $\sqrt{3}a_h \rightarrow c_o$ Δ=-0.5% |
| $(MnNiSi)_{0.62}(FeNiGe)_{0.38}$ [12] | $b_o \rightarrow a_h$ Δ=8.9% | $a_o \rightarrow c_h$ Δ=-11.5% | $c_o \rightarrow \sqrt{3}a_h$ Δ=-0.66% |
| $(MnNiSi)_{0.62}(FeNiGe)_{0.38}$ [13] | $b_o \rightarrow a_h$ Δ=8.8% | $a_o \rightarrow c_h$ Δ=-10.4% | $c_o \rightarrow \sqrt{3}a_h$ Δ=-0.3% |
| $(MnNiSi)_{0.65}(Fe_2Ge)_{0.35}$ [14] | $b_o \rightarrow a_h$ Δ=9.59% | $a_o \rightarrow c_h$ Δ=-10.96 | $c_o \rightarrow \sqrt{3}a_h$ Δ=-0.51% |

*i.e. $a_h \rightarrow b_o$ Δ=-9.40% indicates $\Delta = (b_o - a_h)/a_h = -9.40\%$

Since $[001]_o$ and $[01\bar{1}0]_h$ are the normal directions of $(001)_o$ and $(01\bar{1}0)_h$ surfaces, the habit plane should also be vertical to the two surfaces. We keep $[001]_o$ and $[01\bar{1}0]_h$ vertical to the paper and show the $(001)_o$ and $(01\bar{1}0)_h$ surfaces in Fig. 5a. According to the orientation relationship, $[4\bar{2}\bar{2}3]_h$ is also kept to be parallel to $[120]_o$ in Fig. 5a. For clarity, the atoms in $(01\bar{1}0)_h$ are translucent, while that in $(001)_o$ are solid. The lattice parameters $a_h$, $c_h$, $a_o$ and $b_o$ are also illustrated in this figure. It can be found that the blue and pink rectangles meet at P point. It means that the vector OP doesn't strain and rotate during the transformation. OP is another invariant line. The direction of OP can be generally given by

$$\left[ 2\ \bar{1}\ \bar{1}\ \frac{3\left(x+\frac{c_h}{2}\right)}{c_h} \right] \quad (2)$$

where $x$ is

$$x = \frac{\sin\left(\frac{\pi}{2} - \arctan\frac{a_o}{2b_o}\right)}{\sin\left(\arctan\frac{a_o}{2b_o} - \arctan\frac{c_h}{2a_h}\right)} \left( \sqrt{a_h^2 + \left(\frac{c_h}{2}\right)^2} - \sqrt{b_o^2 + \left(\frac{a_o}{2}\right)^2} \right) \quad (3)$$

(For detail on the calculation, please see Fig. S5 and the corresponding content in supplemental material.) The direction of OP is highly dependent on the lattice parameters of hexagonal and orthorhombic phases. For our sample MnNi$_{0.84}$V$_{0.16}$Ge, OP is along $[2\ \bar{1}\ \bar{1}\ 1.84]_h$.

Based on the observed two invariant lines, the habit plane can be confirmed to be $(\bar{2}113.26)_h$. With the vectors in the habit plane unchanged during the transformation, the hexagonal-orthorhombic transformation can be accomplished by a strain marked in Fig. 5b. This strain consists of two parts. One is vertical to the habit plane, inducing the volume increase during the transformation from hexagonal to orthorhombic structures. The other is a simple shear parallel to the habit plane. This shear rotates the $a_h$ to $b_o$. The angle between $a_h$ and $b_o$ is 5.38° (see Fig. 5a).

To further confirm the obtained habit plane and reveal the origin of this martensitic transformation, we adopt phenomenological crystallographic theory of martensitic transformation (WLR or BM theory) to analyze the relationship between hexagonal and orthorhombic structures. The key of this theory is to find a matrix to express the strain shown in Fig. 5b. According to the WLR theory, this strain, known as the invariant plane strain (denoted as $P_1$), can be represented as follows:

$$P_1 = \varphi P_2 B \tag{4}$$

where matrix B is the deformation matrix $\varepsilon$ as we shown in Formula (1), $P_2$ is the lattice invariant shear, making the deformed bulk ellipsoid tangent to the initial red sphere so that an unstrained plane can be created, and $\varphi$ is a rotation matrix, which can rotate the unstrained vectors back to their initial positions.

For the transformation from hexagonal to orthorhombic, the variation of Z axis is near-zero (see Fig. 1d). So, in the rectangular coordinate shown in Fig. 1d, the blue ellipsoid and red sphere are just tangent to one another. That is to say an additional $P_2$ is not needed for the hexagonal-orthorhombic transformation, or $P_2$ is unit matrix in Formula (4). This situation will greatly simplify the calculation relative to that in TiNi and some other shape memory alloys, in which $P_2$ isn't unit matrix [48]. As we mentioned before, after the deformation $\varepsilon$, the red sphere changes to the blue ellipsoid. The vectors in the blue plane are unstrained, but are rotated by $\theta$. Therefore, by a rotation operation $\varphi$, the blue plane is rotated back to its initial position (red plane) so that the vectors on the red plane are unstrained and unrotated. The red plane is the habit plane according to the phenomenological crystallographic theory. Based on the above-mentioned analysis, matrix $P_1$ is expressed as follows:

$$P_1 = \begin{bmatrix} \cos\theta & -\sin\theta & 0 \\ \sin\theta & \cos\theta & 0 \\ 0 & 0 & 1 \end{bmatrix} \begin{bmatrix} \dfrac{a_o}{c_h} & & \\ & \dfrac{b_o}{a_h} & \\ & & 1 \end{bmatrix} \tag{5}$$

where $\theta$ is the angle between the red and blue plane. $\theta$ can be obtained by solving the following

equation set:

$$X_1^2 + Y_1^2 = 1 \tag{6}$$

$$\frac{X_1^2}{a_o/c_h} + \frac{Y_1^2}{b_o/a_h} = 1 \tag{7}$$

$$\begin{bmatrix} X_1 \\ Y_1 \end{bmatrix} = \varepsilon \begin{bmatrix} X_2 \\ Y_2 \end{bmatrix} \tag{8}$$

$$\cos\theta = \begin{bmatrix} X_2 & Y_2 \end{bmatrix} \begin{bmatrix} X_1 \\ Y_1 \end{bmatrix} \tag{9}$$

Based on the lattice parameters of V-doped MnNiGe shown before, the matrix $P_1$ is given by

$$P_1 = \begin{bmatrix} 1.115 & -0.089 & 0 \\ 0.109 & 0.915 & 0 \\ 0 & 0 & 1 \end{bmatrix} \tag{10}$$

According to $P_1$, the calculated invariant line in XY plane can be obtained to be $[2\bar{1}\bar{1}1.77]_h$ and the direction of $P_1$ is $[211\overline{3.18}]_h$. The calculated habit plane is $(\bar{2}113.40)_h$. The angle between this calculated habit plane and that obtained from TEM measurement is only 1.2°. On the other hand, through the strain $P_1$, $\vec{a_h}$ become $P_1 * \vec{a_h} = \vec{b_o}$. The angle between $\vec{a_h}$ and $\vec{b_o}$ is 5.59°, which is only 0.12° larger than that obtained from TEM measurement. After the strain of $P_1$, $[4\bar{2}\bar{2}3]_h$ become parallel to $[120]_o$. It agrees well with the orientation relationship obtained by TEM measurement. In other words, the calculated result agrees well with that obtained from TEM measurement.

Considering the symmetry of hexagonal structure, we provide all possible habit planes and the corresponding orientation relationships in Table 2. It should be noted that our results don't mean that all the MnMX alloys obey the orientation relationship, such as $[4\bar{2}\bar{2}3]_h // [120]_o$. Whether the two axis are parallel to one another is highly related with the lattice parameter change during the transformation. Because the variation of lattice parameters determines the rotation matrix and the rotation angle θ, which further determines the orientation relationship.

Table 2 All possible habit planes and the corresponding orientation relationships

| Orientation Relationship | Habit Plane |
|---|---|
| $[4\bar{2}\bar{2}3]_h //[120]_o$ & $(01\bar{1}0)_h //(001)_o$ | $(\bar{2}11\,3.26)_h$ |
| $[\bar{4}223]_h //[120]_o$ & $(01\bar{1}0)_h //(001)_o$ | $(2\bar{1}\bar{1}\,3.26)_h$ |
| $[\bar{2}\bar{2}43]_h //[120]_o$ & $(\bar{1}100)_h //(001)_o$ | $(11\bar{2}\,3.26)_h$ |
| $[22\bar{4}3]_h //[120]_o$ & $(\bar{1}100)_h //(001)_o$ | $(\bar{1}\bar{1}2\,3.26)_h$ |
| $[\bar{2}4\bar{2}3]_h //[120]_o$ & $(\bar{1}010)_h //(001)_o$ | $(1\bar{2}1\,3.26)_h$ |
| $[2\bar{4}23]_h //[120]_o$ & $(\bar{1}010)_h //(001)_o$ | $(\bar{1}2\bar{1}\,3.26)_h$ |

## 5. Conclusion

In summery, we take MnNiGe as an example and investigate the crystallography of the martensitic transformation between $Ni_2In$-type hexagonal and TiNiSi-type orthorhombic phases. After tuning the transformation temperature of MnNiGe alloy to be slightly below room temperature, some straight and thin laths in orthorhombic structure are found to distributed in hexagonal matrix. Based on this hexagonal/orthorhombic coexisted sample, HRSTEM measurement is carried out and the results indicate that the hexagonal and orthorhombic structures obey the following orientation relationship: $[4\bar{2}\bar{2}3]_h //[120]_o$ & $(01\bar{1}0)_h //(001)_o$. The TEM results also prove that $[001]_o$ and $[01\bar{1}0]_h$ are in the habit plane. That is why $\sqrt{3}a_h \rightarrow c_o$ is without length change during the transformation. Based on these, the habit plane is confirmed to be $\{\bar{2}11\,3.26\}_h$. Furthermore, WLR theory is adopted to calculate the habit plane. The calculated and measured results agree well with each other. The strain that induces the transformation from hexagonal to orthorhombic structures is also provided here.

**Acknowledgement**

This research is supported by the Fundamental Research Funds for the Central Universities, China (No. 30919012108) and the Natural Science Foundation of China (No: 11974184 ).

**Declaration of Competing Interest**

We declare that we have no financial and personal relationships with other people or organizations that can inappropriately influence our work, there is no professional or other personal interest of any nature or kind in any product, service and/or company that could be construed as influencing the position presented in, or the review of, the manuscript entitled "Crystallography of the martensitic transformation between $Ni_2In$-type hexagonal and TiNiSi-type orthorhombic phases".